\documentclass[english,letterpaper,twocolumn,showpacs,prl,aps]{revtex4}
\usepackage{graphicx}
\usepackage{amssymb}

\makeatletter

\large  

\input epsf

\makeatother

\usepackage{babel}
\makeatother
\begin{document}

\title{On the calculation of the Casimir forces}

\author{Eugene B. Kolomeisky$^{1}$ and Joseph P. Straley,$^{2}$}

\affiliation{$^{1}$Department of Physics, University of Virginia, P. O. Box 400714,
Charlottesville, Virginia 22904-4714, USA\\
$^{{2}}$Department of Physics and Astronomy, University of Kentucky,
Lexington, Kentucky 40506-0055, USA}

\begin{abstract}
Casimir forces are a manifestation of the change in the zero-point energy of the vacuum
caused by the insertion of boundaries.  We show how the Casimir force 
can be computed by consideration of the  
vacuum fluctuations that are \textit{suppressed} by the boundaries,    
and
rederive the scalar Casimir effects for a 
series of geometries.  For the planar case a finite universal force 
is automatically found.  For curved geometries formally divergent 
expressions are encountered which we argue are largely due to 
the divergent self-energy of the boundary contributing to the force.  
This idea is supported by computing the effect for a fixed perimeter 
wedge-arc geometry in two dimensions.
\end{abstract}

\pacs{03.70.+k, 11.10.-z, 11.10.Gh, 42.50.Pq}

\maketitle

Casimir interactions are due to the macroscopic response of the physical vacuum to the introduction of boundaries.  They were first derived as an attractive force between perfectly conductive parallel plates induced by the zero-point motion of 
the electromagnetic field \cite{Casimir}  There is convincing experimental evidence for the reality of these forces \cite{experiment} and a vast body of literature dedicated to various aspects of the phenomenon \cite{CasReviews}.  

The  Casimir interaction $\mathcal{E}$ is the difference between the vacuum 
energy of the system constrained by the boundaries and 
that of free space.  Since boundaries made of real materials are transparent to sufficiently high-energy modes \cite{Casimir}, the high energy spectrum is unaffected by the geometry of the system, and only a finite range of the spectrum need be considered.  However, 
in the theoretical treatments of this effect the vacuum energies are usually calculated from an effective low-energy harmonic field 
theory (such as quantum electrodynamics in the case of the electromagnetic Casimir effect), so that
they are approximated by the sum of zero-point energies of a collection of simple harmonic 
oscillators with a spectrum $\omega = c |\textbf{k}|$ (where $c$ is the speed of 
light).  In this model, the dispersion relation holds for arbitrarily large wave vectors $\textbf{k}$; both the 
"constrained" and "free" vacuum energy densities are ultraviolet divergent; and 
the Casimir interaction is the difference between two infinite quantities.  
The theory resolves this problem by a soft-cutoff modification of the large-\textbf{k} part of the spectrum that leads to a finite vacuum energy.  The result is not very sensitive to the form of the cutoff, so that the Casimir interaction can then be extracted by taking the cutoff to infinity at the end of calculation.  Other approaches to the calculation that make use of analytic continuation \cite{zeta} and dimensional regularization \cite{dim} techniques give the same answer, thus adding to the credibility of the result.

As already 
noted, the divergences are more mathematical artifacts than physical reality.
The important virtue of the model is that for many geometries a finite result is obtained without introducing a cutoff, leading to a finite universal result depending only on $\hbar$, $c$, and macroscopic length scales.   However, this is not always the case: specifically, the divergences occurring for spherical geometry in even space dimensions do not cancel  \cite{Bender,Milton}.  What this means physically represents an open problem;  it seems to imply that in two dimensions
a conducting
ring placed in vacuum is unstable.       

The goal of this Letter is two-fold:  first, we show how the Casimir effect can be computed efficiently by direct consideration of the fluctuation modes that are eliminated by the presence of the boundaries.   Second, we interpret the divergences encountered in the case of curved boundaries as being due to divergent self-energy of the boundary contributing to the Casimir force.  

These ideas will be illustrated by analyzing Casimir effects in a Gaussian field theory with the Euclidian action 
\begin{equation}
\label{action}
S_{E} = {\frac{1}{2}}\int_{0}^{\hbar/T} d\tau d^{d}x \left (c^{-2}({\frac{\partial u}{\partial \tau}})^{2} +
(\nabla u)^{2}\right ),
\end{equation}
where $T$ is the temperature and the real scalar $u$, a function of $d$-dimensional position vector $\textbf{r}$ and imaginary time $\tau$, is periodic on the Matsubara circle, $u(\textbf{r},0) = u(\textbf{r}, \hbar/T)$ \cite{Matsubara}.  The action (\ref{action}) is applicable at energies low compared to some scale $\hbar \omega_{0}$;  corresponding cutoff function will be suppressed and invoked only as needed.

Assume the vacuum is disturbed by the presence of sharp boundaries $D_{i}$ and/or $N_{j}$ of the Dirichlet, $u|_{D_{i}} = 0$, and/or Neumann, $\partial u/\partial n|_{N_{j}} = 0$, type, respectively, where the subscripts $i$ and $j$ label the boundaries and $\partial u/\partial n$ is the normal derivative.   The vacuum fluctuations \textit{eliminated} by these constraints are field configurations which are solutions to the boundary-value problem for the Laplace equation
\begin{equation}
\label{gbvproblem}
(\frac{\partial^{2}}{c^{2}\partial\tau^{2}} + \triangle) u = 0, ~u|_{D_{i}} = f_{i}(\textbf{r},\tau),\frac{\partial u}{\partial n}|_{N_{j}} = g_{j}(\textbf{r}, \tau)
\end{equation} 
where $f_{i}$ and $g_{j}$ are functions defined on the boundaries and playing a role of dynamical variables of our approach.  The Casimir energy is the \textit{negative} of the zero-point energy of the field configurations satisfying (\ref{gbvproblem}) as the latter do not contribute into the energy of the vacuum disturbed by the boundaries.    Since they are static, a simplification is achieved by expanding all the dynamical variables of the problem into Fourier series in imaginary time domain; for example $u(\textbf{r}, \tau) = \sum_{\omega} u_{\omega}(\textbf{r}) \exp i
\omega \tau$
where the Fourier coefficients $u_{\omega}(\textbf{r})$ are solutions to the boundary-value problem for the Helmholtz equation
\begin{equation}
\label{Helmholtz}
(\triangle - \frac{\omega^{2}}{c^{2}})u_{\omega} = 0,~ u_{\omega}|_{D_{i}} = f_{\omega,i}(\textbf{r}),~\frac{\partial u_{\omega}}{\partial n}|_{N_{j}} = g_{\omega,j}(\textbf{r})
\end{equation}  

Calculations of the zero-point energy of the eliminated modes are further  simplified if the identity $(\nabla u)^{2} = div(u\nabla u) - u\triangle u$ is substituted into the action (\ref{action}).  Then the integral of $div(u\nabla u)$ over $d^{d}x$ transforms into a sum of surface integrals.  The remaining integral over $d\tau$ vanishes due to the relation $\triangle u = - \partial^{2} u/c^{2}\partial \tau^{2}$ and the condition of periodicity, $u(\textbf{r},0) = u(\textbf{r}, \hbar/T)$.  As a result we find
\begin{eqnarray}
\label{actionsurfaceintegral}
S_{E}& =& \frac{1}{2} \int_{0}^{\hbar/T} d \tau \sum_{i}\int[u\nabla u]_{i}d\textbf{s}_{i} \nonumber\\& = & \frac{\hbar}{2T}\sum_{\omega,i}\int[u_{\omega}\nabla u_{-\omega}]_{i} d\textbf{s}_{i}
\end{eqnarray}
Here $[\psi]$ stands for the discontinuity of $\psi$ across the boundary, and the summation is performed over all the boundaries.  

The calculation of the zero-point energy of the eliminated field configurations takes advantage of the correspondence between the Feynman path integral for the $d$-dimensional field theory with the action $S_{E}$, $Z = \int Du(\textbf{r},\tau) \exp( - S_{E}[u]/\hbar)$, and a partition function of a $d + 1$-dimensional classical statistical mechanics problem with the Hamiltonian $S_{E}$ at a fictitious temperature which is equal to Planck's constant \cite{Kogut}.  The "free energy" per unit "length" in the imaginary time direction, $-\hbar (\ln Z)/(\hbar /T) = - T \ln Z$, gives the zero-point energy of the field configurations eliminated by the boundaries.   Thus the Casimir energy is $\mathcal{E} = T \ln Z$.  This completes the general formulation of the method which we now illustrate by analyzing zero-temperature Casimir effects in several geometries.  We restrict ourselves to the case of Dirichlet boundaries.

\textbf{Planar geometry.}  Consider three Dirichlet planes at $z = 0$, $z = a$, and $z = L$, where $z$ is one of the axes of the $d$-dimensional rectangular coordinate system and $0 < a < L$.  The outer boundaries are fixed in place so that there is no need to look beyond them.  We are interested in the Casimir pressure exerted on the middle partition at $z = a$.  Since the space is uniform relative to translations parallel to the boundaries, the field $u_{\omega}(\textbf{r})$ is expanded into a Fourier series $u_{\omega}(\textbf{r}) = \sum_{\small\textbf{q}} u_{\omega \small\textbf{q}}(z) \exp i \textbf{q} \textbf{r}_{\perp}$ where $\textbf{r}_{\perp}$ is the position vector perpendicular to the $z$ axis.  Then the boundary-value problem (\ref{Helmholtz}) for the Fourier coefficients $u_{\omega \small\textbf{q}}(z)$ becomes
\begin{equation}
\label{bvplane}
(\frac{d^{2}}{dz^{2}} - q^{2} - \frac{\omega^{2}}{c^{2}})u_{\omega \small\textbf{q}} = 0, ~u_{\omega \small\textbf{q}}|_{0,L} = 0,~u_{\omega \small\textbf{q}}|_{a} = f_{\omega \small\textbf{q}}
\end{equation}
The solution to (\ref{bvplane}) is
\begin{eqnarray}
\label{planesolution}
u_{\omega \small\textbf{q}}(z)& = &f_{\omega \small\textbf{q}}\frac{\sinh (|\kappa| z)}{\sinh (|\kappa| a)}, ~~~~~~0 \leqslant z \leqslant a, \kappa^{2} = q^{2} + \frac{\omega^{2}}{c^{2}}\nonumber\\
u_{\omega \small\textbf{q}}(z)& = &f_{\omega \small\textbf{q}}\frac{\sinh (|\kappa| (L - z))}{\sinh (|\kappa| (L - a))}, ~a < z \leqslant L
\end{eqnarray}
Substituting this in Eq.(\ref{actionsurfaceintegral}) we see that only the partition at $z = a$ contributes into the action $S_{E}$ with the result
\begin{equation}
\label{actionplanar}
S_{E} = \frac{\hbar \mathcal{A}}{2T} \sum_{\omega, \small\textbf{q}} |\kappa| (\coth (|\kappa| a) + \coth (|\kappa|(L - a)))|f_{\omega \small\textbf{q}}|^{2}
\end{equation}
where $\mathcal{A}$ is the macroscopic $(d - 1)$-dimensional area of the boundary.  Then up to an $a$-independent constant the Casimir energy per unit area is  given by
\begin{eqnarray}
\label{Casimirenergyplanar}
&&\frac{\mathcal{E}}{\mathcal{A}} = -\frac{T}{2\mathcal{A}} \sum_{\omega, \small\textbf{q}} \ln \left (\coth (|\kappa| a) + \coth (|\kappa|(L - a))\right )\nonumber\\
& \rightarrow & -\frac{\hbar}{2} \int \frac{d\omega d^{d - 1}q}{(2\pi)^{d}} \ln \left (\coth (|\kappa | a) + \coth (|\kappa |(L - a))\right )\nonumber\\
& = & - \frac{\hbar c K_{d}}{2} \int _{0}^{\infty}\kappa^{d-1} d\kappa \ln \left (\coth (\kappa a) + \coth (\kappa (L - a))\right )\nonumber\\
\end{eqnarray}
where in taking the $T = 0$ and macroscopic limits we used the rules $\sum_{\omega} \rightarrow (\hbar/T) \int d\omega/2\pi$, $\sum_{\small\textbf{q}} \rightarrow \mathcal{A} \int d^{d-1}q/(2\pi)^{d-1}$, respectively.  The parameter $K_{d}$ in the third representation is the surface area of a $d$-dimensional unit sphere, $2\pi^{d/2}/\Gamma(d/2)$, divided by $(2\pi)^{d}$.  The Casimir pressure on the boundary, $\mathcal{P} = - \partial (\mathcal{E}/\mathcal{A})/\partial a$ can be found in closed form
\begin{equation}
\label{planarpressure}
\mathcal{P} =  \frac{d \Gamma(\frac{d+1}{2}) \zeta(d + 1)} {(4\pi)^{\frac{d+1}{2}}}\hbar c\left (\frac{1}{(L - a)^{d + 1}} - \frac{1}{a^{d+1}}\right )
 \end{equation}   
 where $\Gamma(x)$ and $\zeta(x)$ are Euler's and Riemann's gamma and zeta functions, respectively \cite{AS}.  In arriving at (\ref{planarpressure}) we used the gamma function duplication formula \cite{AS} and the value of the integral $\int_{0}^{\infty}x^{d}(\coth x - 1)dx = 2^{-d}\Gamma(d + 1) \zeta(d + 1)$ \cite{integral}.

We see that the partition at $z = a$ is attracted to the closest outer boundary; specifically, in one dimension Eq.(\ref{planarpressure}) reduces to the well-known result \cite{Boyer}.  Since the outer boundaries impose the Dirichlet boundary conditions we can imagine joining them together.  Then Eq.(\ref{planarpressure}) describes Casimir interaction between two boundaries;  taking the $L \rightarrow \infty $ limit we then reproduce the result of Ambj\o rn and Wolfram \cite{dim}.

\textbf{Circular geometry.}  Consider a Dirichlet circle of radius $a$ in two spatial dimensions.  The boundary-value problem (\ref{Helmholtz}) for this geometry becomes
\begin{equation}
\label{bvcircle}
\left (\frac{1}{\rho}\frac{\partial}{\partial \rho}(\rho \frac{\partial}{\partial \rho}) + \frac{1}{\rho^{2}}\frac{\partial^{2}}{\partial\varphi^{2}} - \frac{\omega^{2}}{c^{2}}\right )u_{\omega} = 0,u_{\omega}|_{a} = f_{\omega}(\varphi)
\end{equation}
where $\rho$ and $\varphi$ are the polar coordinates.  Seeking the particular solution in the form, $u_{\omega}(\rho, \varphi) = R_{\omega}(\rho)\exp i n\varphi$, where $n$ is an arbitrary integer, we find that the radial function $R_{\omega}(\rho)$ satisfies  the equation \begin{equation}
\label{Bessel}
\frac{d^{2}R_{\omega}}{d\rho^{2}} + \frac{1}{\rho}\frac{dR_{\omega}}{d\rho} - (\frac{\omega^{2}}{c^{2}} + \frac{n^{2}}{\rho^{2}})R_{\omega} = 0
\end{equation}
whose linearly-independent solutions are modified Bessel functions $I_{n}(|\omega|\rho/c)$ and $K_{n}(|\omega|\rho/c)$ \cite{AS}.  Thus the general solution to the boundary-value problem (\ref{bvcircle}) finite at $\rho = 0$ and decaying as $\rho \rightarrow \infty$  is
\begin{eqnarray}
\label{circle solution}
u_{\omega}& = &\sum_{n=-\infty}^{\infty}\frac{I_{n}(|\omega |\rho/c)}{I_{n}(|\omega |a/c)} f_{\omega n} \exp in\varphi, ~~~\rho \leqslant a \nonumber\\
u_{\omega}& = &\sum_{n=-\infty}^{\infty}\frac{K_{n}(|\omega |\rho/c)}{K_{n}(|\omega |a/c)} f_{\omega n} \exp in\varphi, ~~~\rho > a
\end{eqnarray}
At the circle $\rho = a$ both of these reduce to $f_{\omega}(\varphi) = \sum_{n = -\infty}^{\infty} f_{\omega n}\exp i n\varphi$.  Substituting the solution (\ref{circle solution}) in Eq.(\ref{actionsurfaceintegral}), using the Wronskian $K_{n}(z)I_{n}^{'}(z) - I_{n}(z)K_{n}^{'}(z) = 1/z$ \cite{AS}, and performing angular integration we find
\begin{equation}
\label{circleaction}
S_{E} = \frac{\pi \hbar}{T} \sum_{\omega,n}\frac{|f_{\omega n}|^{2}}{I_{n}(|\omega|a/c) K_{n}(|\omega| a/c)}
\end{equation}
The Casimir energy $\mathcal{E}$ and the force $\mathcal{F} = - \partial \mathcal{E}/\partial a$  are then given by
\begin{equation}
\label{Casenergycircle}
\mathcal{E} = \frac{\hbar c}{2\pi}\sum_{n = -\infty}^{\infty} \int_{0}^{\infty} d\kappa \ln\left (const I_{n}(\kappa a)K_{n}(\kappa a)\right )
\end{equation}
\begin{equation}
\label{Casforcecircle}
\mathcal{F} = -\frac{\hbar c}{2\pi a^{2}} \sum_{n = -\infty}^{\infty} \int_{0}^{\infty}zdz \frac{d}{dz}\left(\ln(I_{n}(z)K_{n}(z))\right )
\end{equation}
where the \textit{const} includes all $a$-independent parameters not contributing into the force.  The results (\ref{Casenergycircle}) and (\ref{Casforcecircle}) are due to Sen \cite{Sen} (see also Refs.  \cite{Ng,Bender}).

We have verified that the Casimir effects for spherical \cite{Bender} and cylindrical geometries \cite{Nesterenko} and the cases of the Neumann and mixed boundary conditions can be treated similarly; unfortunately, we have no space here to enter into details of the analysis. 

\textbf{Divergences and their interpretation.}  Although Eq.(\ref{planarpressure}) solves the planar version of the problem predicting finite universal Casimir pressure,  its circular (and any curvilinear) counterpart  (\ref{Casforcecircle}) is divergent.  Various researchers dealt with this issue differently.  Sen \cite{Sen} viewed the action (\ref{action}) as an effective low-energy theory which must be supplemented by a cutoff function.  Introducing this into the integrand of Eq.(\ref{Casenergycircle}) removes the divergence, thus predicting finite non-universal effect.  Other researchers attempted to remove the divergences by a variety of techniques \cite{Bender} but a divergence was still found in the two-dimensional circular case.  

In order to understand the difference between planar and curved geometries we notice that the Casimir force is the change of the energy upon infinitesimal displacement of the boundary.  A geometrically sharp boundary possesses a formally divergent energy per unit area.  This divergent self-energy does not contribute to the Casimir force in the planar case as the overall area remains fixed as the boundary is displaced.  However this is not the case for curved boundaries.  Indeed a change of the radius of a circle implies a change of the perimeter and as a result the divergent self-energy will contribute into the force.  We argue that this is the underlying reason behind divergences encountered for curved geometries.  In order to test this idea we need to look at a curved geometry where the area of the boundaries can be kept fixed. 
\begin{figure}
\includegraphics[width=1.0\columnwidth,keepaspectratio]{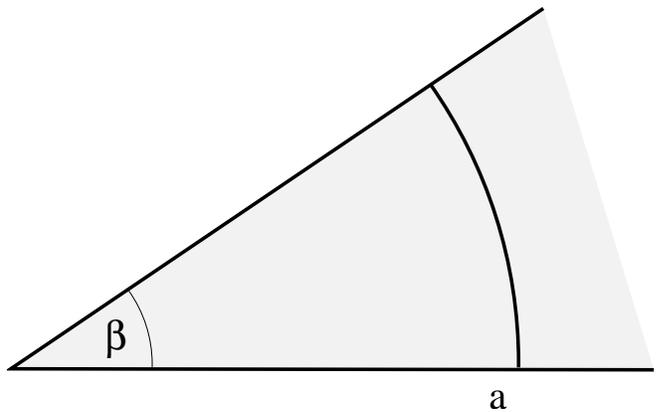} 
\caption{Wedge of opening angle $\beta$ with superimposed arc of radius $a$ in two dimensions.}
\end{figure}

\textbf{Wedge-arc geometry in two dimensions.}  Consider a wedge of opening angle $\beta$ with superimposed arc of radius $a$, Fig. 1.  This is the geometry where both a Casimir force and torque occur \cite{note}.  Additionally, if the variations of $\beta$ and $a$ are constrained by the condition of fixed arc length $l = \beta a$, the net perimeter of the infinite wedge edges and the arc are fixed.  The Casimir interaction here can be inferred from the results for the circular geometry since the boundary-value problem we need to solve is closely related to (\ref{bvcircle}).  The  difference is that we seek a solution inside the Dirichlet wedge, $0 \leqslant \varphi \leqslant \beta$ thus implying $u_{\omega}(\rho,\varphi) = R_{\omega}(\rho) \sin(\pi n \varphi/\beta)$, $n = 1, 2, ...$ for the particular solution. The radial function $R_{\omega}(\rho)$ satisfies the same Eq.(\ref{Bessel}) with $n$ being  replaced by $\pi n/\beta$.  As a result the solution in question can be obtained from Eq.(\ref{circle solution}) by replacing the order $n$ of the Bessel functions with $\pi n/\beta$, the angular function $\exp i n\varphi$ with $\sin(\pi n\varphi/\beta)$, and restricting the summation over $n$ from $1$ to infinity.  The calculation of the Casimir energy is similar to that for the circular geometry with the result
\begin{equation}
\label{casenergywedge}
\mathcal{E} = \frac{\hbar c}{2\pi a}\sum_{n = 1}^{\infty} \int_{0}^{\infty} dz \ln\left (\frac{const}{\beta} I_{\frac{\pi n}{\beta}}(z)K_{\frac{\pi n}{\beta}}(z)\right )
\end{equation} 
where the $const$ is both $a$ and $\beta$ independent.   We then employ the uniform asymptotic $n \gg 1$ expansion of Debye \cite{AS,Schwinger}:
\begin{eqnarray}
\label{Debye}
2n(1&+&x^{2})^{1/2}I_{n}(nx)K_{n}(nx) = 1 + \frac{1}{n^{2}}(\frac{0.125}{1+x^{2}}\nonumber\\& -& \frac{0.75}{(1+x^{2})^{2}} + \frac{0.625}{(1+x^{2})^{3}}) + \mathcal{O}(\frac{1}{n^{3}})
\end{eqnarray}  
which can be used to evaluate the energy (\ref{casenergywedge}) in the $\beta \ll 1$ limit.  To leading order we find
\begin{eqnarray}
\label{firstorder}
\mathcal{E}^{(1)}& =& \frac{\hbar c}{2\pi a}\sum_{n=1}^{\infty}\int_{0}^{\infty}dz\ln \left (\frac{const}{2\pi n(1+(\beta z/\pi n)^{2})^{1/2}}\right )\nonumber\\
&=&\frac{\hbar c}{2\beta a} \sum_{n=1}^{\infty}n\int_{0}^{\infty}dt \ln\left (\frac{const}{2\pi n (1+t^{2})^{1/2}}\right )
\end{eqnarray}
Although this expression is formally divergent, it depends on $a$ and $\beta$ only through the fixed arc length $l = \beta a$.  Thus (\ref{firstorder}) is a constant which can be subtracted from (\ref{casenergywedge}).  As a result we find for presumably universal part of the Casimir energy $\mathcal{U} = \mathcal{E} - \mathcal{E}^{(1)}$:   
\begin{equation}
\label{wedgeenergyuniversal}
\mathcal{U} = \frac{\hbar c}{2l}\sum_{n=1}^{\infty}n\int_{0}^{\infty} dt \ln (2\frac{\pi n}{\beta}(1+t^{2})^{\frac{1}{2}}I_{\frac{\pi n}{\beta}}(\frac{\pi n}{\beta}t) K_{\frac{\pi n}{\beta}}(\frac{\pi n}{\beta}t))
\end{equation}
Using the expansion (\ref{Debye}) again we evaluate the energy to the next order (\ref{wedgeenergyuniversal}) in the $\beta \ll 1$ limit with the result
\begin{equation}   
\label{wedgeenergydivergent}
\mathcal{U} = - \frac{\hbar c \beta^{2}}{256\pi l}\sum_{n=1}^{\infty}n^{-1}, ~~~\beta \ll 1, 
\end{equation}
which is marginally divergent.  This means there is a weak dependence on the cutoff frequency $\omega_{0}$.  The latter can be recovered by employing the equality $\sum_{n=1}^{N}n^{-1} = C + \ln N + \epsilon_{N}$ where $C$ is Euler's constant and $\epsilon_{N} \rightarrow 0$ as $N \rightarrow \infty$ \cite{AS}.  The parameter $N$ is estimated by recalling that a cutoff function $F(cz/\omega_{0}a)$ is suppressed from the integrand of Eq.(\ref{casenergywedge}).  In terms of the variable $t$ of Eq.(\ref{wedgeenergyuniversal}) this becomes $F(\pi nct/\omega_{0}l)$ which removes the divergence in (\ref{wedgeenergydivergent}) by effectively ending the sum of $1/n$ at $n = N$ such as $\pi nct/\omega_{0}l \approx 1$ and $t \approx 1$.  Thus $N \approx \omega_{0}l/c$.  As a result we find that with logarithmic accuracy
\begin{equation}
\label{final}
\mathcal{U} = - \frac{\hbar c \beta^{2}}{256\pi l}\ln\frac{\omega_{0}l}{c}, ~~~\beta \ll 1, ~~~\frac{\omega_{0}l}{c} \gg 1
\end{equation}
This result supports our idea that the divergences are due to the boundary self-energy because the dependence on the cutoff frequency $\omega_{0}$ is logarithmically weak and the amplitude of the logarithmic dependence is universal.  Sen's result for the circle \cite{Sen} also has a term logarithmically dependent on the cutoff frequency $\omega_{0}$.  

Our result (\ref{final}) means that for fixed arc length $l$ the zero-point motion induces a \textit{widening} torque $-\partial \mathcal{U}/\partial \beta$ which is not unexpected on the physical grounds.  The accuracy of the Debye expansion (\ref{Debye}), implies that Eq.(\ref{final}) approximately captures the whole $0\leqslant \beta \leqslant 2\pi$ range.  

To summarize, we have demonstrated how Casimir effects caused by sharp boundaries can be efficiently computed by focusing on the quantum fluctuations eliminated by these boundaries.  The applicability of this method is not limited to the scalar field theory (\ref{action}), Dirichlet boundaries, zero-temperature limit or to the geometries we have considered.  

Additionally, we argued that the formally divergent Casimir forces encountered in the presence of curved boundaries are due to divergent self-energy contributions, and supported this idea by an explicit calculation.  More work is needed to further test this hypothesis.   Our analysis also supports Sen's  viewpoint \cite{Sen} that the Casimir effect in a Dirichlet ring in two dimensions is finite and non-universal with the cutoff frequency $\omega_{0}$ supplied by the properties of the material the boundary is made of. 

This work was supported by the Thomas F. Jeffress and Kate Miller Jeffress Memorial Trust.


\begin{thebibliography}{10}
\bibitem{note}  Wedge geometries have been studied in the past \cite{Milton} but not the one shown in Fig.1.
\end{thebibliography}
\end{document}